\documentclass[adraft,copyright,creativecommons]{eptcs}
\usepackage{breakurl}             
\usepackage{underscore}           

\usepackage{amsmath, amsthm, amssymb, amsfonts}
\usepackage{bussproofs}

\title{Towards a Logic for Reasoning About LF Specifications}
\author{Mary Southern \qquad\qquad Gopalan Nadathur
\institute{University of Minnesota\\Minneapolis Minnesota, USA}
\email{south163@umn.edu \qquad\qquad ngopalan@umn.edu}
}
\def\titlerunning{Towards a Logic for Reasoning About LF Specifications}
\def\authorrunning{M. Southern \& G. Nadathur}

\newcommand{\ignore}[1]{}

\newcommand{\lfprove}[3]{#2 \vdash_{#1} #3}
\newcommand{\lfder}[2]{#1 \vdash #2}

\newcommand{\seq}[4]{#2;#1;#3\longrightarrow #4}
\newcommand{\seqsans}[2]{#1\longrightarrow #2}
\newcommand{\subs}[2]{#1(#2)}

\newcommand{\typedpi}[3]{\Pi {#1}{:}{#2}.{\mkern 3mu} #3}
\newcommand{\type}{\mbox{\it Type}}
\newcommand{\imp}[2]{#1\rightarrow #2}
\newcommand{\app}{\;}

\newcommand{\eq}{{\it eq}}
\newcommand{\refl}{{\it refl}}
\newcommand{\sarr}{{\it arr}}
\newcommand{\sapp}{{\it app}}
\newcommand{\slam}{{\it lam}}
\newcommand{\sty}{{\it ty}}
\newcommand{\stm}{{\it tm}}
\newcommand{\of}{{\it of}}
\newcommand{\ofapp}{{\it of\_app}}
\newcommand{\oflam}{{\it of\_lam}}

\begin{document}
\maketitle

\begin{abstract}
We describe the development of a logic for reasoning about
specifications in the Edinburgh Logical Framework (LF).  
In this logic, typing judgments in LF serve as atomic formulas,
and quantification is permitted over contexts and terms that might
appear in them.
Further, contexts, which constitute type assignments to uniquely named
variables that are modelled using the technical device of
\emph{nominal constants}, can be characterized via an inductive
description of their structure.
We present a semantics for such formulas and then consider the task
of proving them.
Towards this end, we restrict the collection of formulas we
consider so as to ensure that they have normal forms upon which proof
rules may be based.
We then specifically discuss a proof rule that provides the basis
for case analysis over LF typing judgments; this rule is the most
complex and innovative one in the collection. 
We illustrate the proof system through an example. 
Finally, we discuss ongoing work and we relate our project to existing
systems that have a similar goal.

\end{abstract}

\section{Introduction}
This paper concerns the development of a system for reasoning about
specifications written in LF~\cite{harper93jacm}.
We aim to do this by designing a logic which allows us to state and
prove relationships between the derivability of different typing
judgments in LF.
Not unexpectedly, the atomic formulas of the logic will be typing
judgments in LF. 
Quantification is permitted over the contexts and terms which appear
in these judgments.
To facilitate reasoning about formulas involving contexts we characterize
them via inductive definitions of their structures.
This style of characterization is motivated by context definitions
that are used in Abella~\cite{baelde14jfr} and has significant
similarities to the  regular worlds descriptions used in the 
Twelf system~\cite{Pfenning02guide}.
In essence, they capture the way contexts change dynamically in LF typing
derivations. 
We provide a semantics for our formulas that is based on the
derivability of typing judgments in LF.
We then outline a proof system for deriving formulas in our logic.
A key rule in this logic, the only one we discuss explicitly here, is
that of case analysis  applied to LF typing judgments that appear as
assumptions.
A problem that we must solve in developing an effective proof system
is that of dealing with quantification over term variables whose types
are eventually constrained by atomic judgments that appear within the
scope of the quantifier. 

The rest of the paper is laid out as follows.
In Section~\ref{sec:formulas} we present the formulas of our logic.
In Section~\ref{sec:rules} we describe a semantics for these formulas
and outline a calculus that can be used to prove them. 
We illustrate the proof system we propose through an example in 
Section~\ref{sec:example}.
In Section~\ref{sec:furtherwork} we discuss ongoing work that includes
proving meta-theoretic properties of our proof system. 
Section~\ref{sec:conclusion} concludes the paper by contrasting our
work with existing approaches that have a similar goal.

\section{Formulas Over LF Typing Judgements and their Semantics}
\label{sec:formulas}

The formulas in our logic build on typing judgments in LF. 
We assume familiarity with the presentation of LF
in~\cite{harper93jacm} and we recall only a few notions that are
needed to set up the discussion here. 
There are three categories of expressions in LF: kinds, type families
and objects.
Objects are classified by type families and type families are
classified by kinds. 
The construction of type families and objects is parameterized by a
signature that, in the intended use, specifies an object system.
For example, if our focus is on typing judgments concerning the simply
typed $\lambda$-calculus, then we might use the signature shown in
Figure~\ref{fig:stlc-in-lf}. 
The core of LF consists of rules for
determining whether expressions of varied kinds are well-formed or
well-typed. 
Our focus is on judgments that take the form
$\lfprove{\Sigma}{\Gamma}{M:A}$, asserting that $M$ is a well-formed
object that inhabits the (well-formed) type family $A$ given the
signature $\Sigma$ and a context $\Gamma$ that assigns types to
variables that might appear free in $M$ and $A$.
Intrinsic to our applications is the equality of expressions based on
$\lambda$-convertibility. 
The typing rules guarantee the existence of canonical froms under
$\lambda$-conversion for well-formed expressions.
However, weaker conditions may also suffice to provide such a
guarantee, an observation that will be useful in the next section.

\begin{figure}[h]
\begin{center}
\begin{tabular}{c}

\begin{tabular}{lll}
$\sty : \type\qquad$ & $\stm : \type\qquad$ & $\of : \imp{\stm}{\imp{\sty}{\type}}$
\end{tabular}\\[10pt]

\begin{tabular}{ccc}
     $\sarr : \imp{\sty}{\imp{\sty}{\sty}}\qquad$ 
   & $\sapp : \imp{\stm}{\imp{\stm}{\stm}}\qquad$
     $\slam : \imp{\sty}{\imp{(\imp{\stm}{\stm})}{\stm}}\qquad$ 
\end{tabular}\\[10pt] 

\begin{tabular}{l}
$\ofapp :\typedpi{M_1}{\stm}
                 {\typedpi{M_2}{\stm}
                          {\typedpi{{\it Ty}_1}{\sty}
                                   {\typedpi{{\it Ty}_2}{\sty}}}}$\\
\qquad$\of\app M_1\app (\sarr\app {\it Ty}_1\app {\it Ty}_2) \rightarrow 
            \of\app M_2\app {\it Ty}_1 \rightarrow 
            \of\app (\sapp\app M_1\app M_2)\app {\it Ty}_2$\\[5pt]
$\oflam  : \typedpi{{\it Ty}_1}{\sty}
                   {\typedpi{{\it Ty}_2}{\sty}
                            {\typedpi{M}{\imp{\stm}{\stm}}{}}}$\\
\qquad$(\typedpi{x}{\stm}
                { \of\app x\app {\it Ty}_1 \rightarrow 
                      \of\app (M\app x)\app {\it Ty_2} \rightarrow 
                      \of \app (\slam \app {\it Ty}_1 \app  M) 
                               (\sarr\app {\it Ty}_1\app {\it Ty}_2)})$
\end{tabular}
\end{tabular}
\end{center}
\caption{An LF signature for the simply typed $\lambda$-calculus} 
\label{fig:stlc-in-lf}
\end{figure}


Our formulas will permit quantification over LF contexts. 
Contexts can comprise a generic part and a part that explicitly
assigns types to variables.
%
To model the fact that explicit assignments have to be to distinct
variables, we use the technical device of \emph{nominal
  constants}~\cite{gacek11ic,tiu06lfmtp}. 
These constants are represented using the token $n$ possibly with
subscripts below. 
The syntax of context expressions, then, is the following:
\begin{align*}
G \quad &::= \quad \Gamma\ |\ \cdot\ |\ G,n:A
\end{align*} 
The symbol $\Gamma$ denotes the category of variables that range over
contexts here.  
In typical reasoning scenarios, the contexts that appear in formulas
will have to be constrained in order to be able to derive contentful
properties. 
Taking inspiration from Twelf~\cite{Pfenning02guide,schurmann00phd},
we provide a means for articulating such structure using \emph{blocks}
that describe the types associated with a finite sequence of context 
variables; the actual context is then understood to be a repetition of
such blocks. 
This idea is formally embodied in declarations given by the following
syntax rules. 
\begin{align*}
Block \quad &::=\quad (x_1:A_1,\ldots,x_n:A_n)[y_1:B_1,\ldots,y_m:B_m]\\
Blocks \quad &::=\quad Block\ |\ Blocks,Block\\
Context\ Schema \quad &::=\quad id = Blocks
\end{align*}
According to these rules, a \emph{context schema} associates a name
with a collection of block descriptions. 
The sequence of declarations $n_1:C_1,\ldots,n_m:C_m$ instantiates the
block 
$(x_1:A_1,\ldots,x_n:A_n)[y_1:B_1,\ldots,y_m:B_m]$ if, for $1 \leq i
\leq m$, $C_i =
B_i[t_1/x_1,\ldots,t_n/x_n,n_1/y_1,\ldots,n_{i-1}/y_{i-1}]$. 
A context satisfies a context schema given by $ctx$ if it comprises
a collection of declaration sequences that instantiate blocks
corresponding to $ctx$.

The collection of formulas determining the logic is given by the
following syntax rule:
\[
F ::= \{\lfder{G}{M : A}\}\ |\ \forall X.F\ |\ \exists X.F\ |\ \Pi_{id}\Gamma.F
      \ |\ F_1\wedge F_2\ |\ F_1\vee F_2\ |\ F_1\supset F_2\ |\ \top\ |\ \bot
\]
The symbol $X$ represents LF object level variables. 
The key to understanding the meaning of these formulas is
understanding the interpretation of formulas of the form
$\{\lfder{G}{M : A}\}$ that do not contain any object or context level variables.
Such a judgment is deemed to be true exactly when 
$G \vdash_\Sigma M : A$ is derivable in LF; $\Sigma$ represents the
implicitly assumed signature here.
The formula $\Pi_{id} \Gamma. F$ holds just in the case that
$F[G/\Gamma]$ holds for every valid LF context $G$ that satisfies the 
context schema $id$.
The formula $\forall X.F$ holds exactly when $F[t/X]$ holds
for every closed expression $t$; observe that instantiations for the
quantified variable are not constrained by types but typing
constraints can become relevant to the truth of the overall formula
depending on where the variable $X$ occurs in $F$. 
The formula $\exists X. F$ is to be understood similarly and the 
propositional connectives have the obvious compositional
interpretation. 

We illustrate the way in which our formulas can be used to encode
properties of object systems by considering the statement of
uniqueness of typing for the simply typed $\lambda$-calculus.
We formulate this property more generally as one about terms that
might have free variables in them for which the type assignments are
provided in an LF context. 
The structure of such a context is given by the following context
schema definition: $tyctx = (T:ty).[x:tm,y:\of~x~T].$
The property of interest is then expressed via the following formula:
\[ \Pi_{tyctx}\Gamma.\forall E.\forall {\it Ty}_1.\forall {\it Ty}_2.\forall D_1.\forall D_2.
\{\lfder{\Gamma}{D_1:\of~E~{\it Ty}_1}\}\supset
\{\lfder{\Gamma}{D_2:\of~E~{\it Ty}_2}\}\supset 
\exists D_3.\{\lfder{\cdot}{D_3:eq\app {\it Ty}_1\app {\it Ty}_2}\}. \]
The signature that implicitly governs this statement is that shown in
Figure~\ref{fig:stlc-in-lf} augmented with the following additional
declarations: $\eq : ty \rightarrow ty \rightarrow \type$ and
$\refl : \Pi T:ty. \eq\app T\app T$. 
\section{Proving Formulas in the Logic}
\label{sec:rules}

Our objective is to develop a calculus that can be used to prove
formulas of the kind described in the previous section.
The calculus will allow us to prove sequents whose form we present in
the first subsection below. 
Our proof system will build in $\lambda$-conversion over object level
expressions in LF.
It will do this by basing the rules on normal forms and assuming that
expressions have been converted into such a form prior to considering
rule applications. 
We impose a restriction on formulas towards making this
approach feasible in the second subsection.
In the third subsection we discuss the most intricate and interesting
of our proof rules, the one for case analysis applied to atomic
judgments that encode LF derivability in formulas. 
Space restrictions prevent us from presenting the other rules.
However, these have a structure that can be easily guessed and we will
use their informal understanding in a subsequent illustration.

\subsection{The Structure of Sequents}

The sequents used in the proof system comprise four parts: a
collection of eigenvariables that range over terms, a set of 
partially elaborated contexts, a collection of assumption formulas and
a conclusion formula. Denoting these respective parts by ${\cal C}$,
$\Psi$, $\Delta$ and $F$, sequents will be written as
$\seq{\mathcal{C}}{\Psi}{\Delta}{F}$.   
The goal of showing that $F$ is a theorem then amounts to trying to
construct a proof for the sequent $\seq{\cdot}{\cdot}{\cdot}{F}$,
where $\cdot$ represents the empty collection. 

The only component amongst those constituting a sequent that needs
further elaboration is the collection of partially elaborated
contexts.
These consist of elements of the form $\Gamma : id\lbrack {\cal
B}\rbrack$ where $id$ is the name for a context schema and ${\cal B}$
is a finite sequence $b_1,\ldots,b_n$ of block instances.
Each such declaration in $\cal C$ is said to introduce the variable
$\Gamma$ and to give it the type $id\lbrack {\cal B}\rbrack$.
The type in such a declaration must satisfy the condition that each
$b_i$ is an instantiation of a block in the context schema associated
with $id$. 
Note that variables may appear in the terms in $b_i$ but such variables
must be contained in the eigenvariable context as we clarify presently.
A variable $\Gamma$ that is typed in this fashion
is intended to be read schematically as
$\Gamma_1,b_1,\ldots,\Gamma_m,b_m,\Gamma_{m+1}$ where each $\Gamma_i$
represents an as yet unelaborated part of the context.
An LF context is a valid instance for $\Gamma$ as per its declaration and under
a closed substitution $\theta$ for the variables appearing in 
$b_1,\ldots,b_m$ if it is a valid LF context that is obtained by
replacing (a)~each $\Gamma_i$ in the schematic view by LF contexts
that satisfy the context schema given by $id$ and (b)~each type
assignment $n:A$ in each $b_i$ for $1 \leq i \leq m$ by $n
: \subs{\theta}{A}$.
By a closed substitution we mean one whose range consists of
expressions not containing nominal constants or free variables.
Further, we use the notation $\subs{\theta}{U}$ here and below to
denote the result of applying $\theta$ in a capture-avoiding way to a
type- or object-level expression $U$. 

We require all the object-level free variables appearing in
$\mathcal{C}$ in the sequent $\seq{\mathcal{C}}{\Psi}{\Delta}{F}$ to
be included in $\Psi$. 
We further require all the free variables ranging over contexts
and terms appearing in $\Delta$ or $F$  to be included in
$\mathcal{C}$ and $\Psi$, respectively.
We then associate a semantics with a sequent of this form as follows.
First, if $\Delta$ and $F$ do not contain any free variables, then 
$\seqsans{\Delta}{F}$ is said to be valid if all the assumption formulas
in $\Delta$ holding implies that $F$ holds.
Now, given $\seq{\mathcal{C}}{\Psi}{\Delta}{F}$ and a closed
substitution $\theta$ for the variables in $\Psi$, we can generate a
variable-free structure of this kind by applying $\theta$ to each
formula in $\seqsans{\Delta}{F}$ and by further replacing each
context-level free variable by an LF context that is valid for that
variable as per its definition in $\mathcal{C}$ under $\theta$.
The sequent $\seq{\mathcal{C}}{\Psi}{\Delta}{F}$ is said to be valid
if it is the case that every structure we can generate in this way
for every substitution $\theta$ for the variables $\Psi$ is valid.

\subsection{A Restriction on Formulas}

We have permitted an untyped form of quantification over terms in
formulas. 
This does not prevent us from providing a reasonable semantics for
formulas because the LF typing judgments that appear as constituents
of these formulas will eventually impose relevant typing constraints.
However, it does pose a problem from the perspective of constructing
proofs for formulas in a situation where we want to build in equality
based on $\lambda$-conversion.
One important issue is the following: we would like our proof rules to
be based on  $\lambda$-normal forms for formulas but, in the absence
of any form of typing, such forms are not guaranteed to exist.
Rectifying this situation by imposing an LF-style typing on quantified
variables is not a possibility.
The reason for this is that such types can encode dependencies on
contexts whose expression should be left to the atomic constituents of
a formula and must not be imposed by the enclosing quantifier.
An alternative solution is to require formulas to be typeable in a less
restrictive fashion that circumvents this issue but that,
nevertheless, guarantees the existence of $\lambda$-normal forms.

We adopt the latter solution here.
In particular, we adapt the typing rules of LF to use a weakened
notion of type matching that ignores dependencies in types: for any
type-level constant $a$, the types $(a\app t_1\app \ldots\app t_n)$
and $(a\app s_1\app \ldots\app s_n)$ match regardless of the values of
the arguments provided to $a$ in the two cases; and the types
$\typedpi{x}{A_1}{B_1}$ and $\typedpi{x}{A_2}{B_2}$ match if $A_1$ and
$A_2$ match and so also do $B_1$ and $B_2$.
We say that a context is weakly valid if it is valid under the
relevant LF typing rules with the exception that this notion of
matching of types is used.
Similarly, we say that $M$ is weakly well-typed at type $A$ in the
(explicitly provided) context $\Gamma$ if
$\lfprove{\Sigma}{\Gamma}{M:A}$ is derivable using the weaker notion
of type matching. 
Now, an atomic formula in the logic has the form
$\{\Gamma,n_1:A_1,\ldots,n_m : A_m \vdash M : A\}$, where $\Gamma$ is
a variable representing an as yet unspecified part of the context.
Such a formula is deemed to be well-typed under an assignment
$\Xi$ of types to the object-level variables appearing in it if $M$ is
weakly well-typed at type $A$ in the context $\Xi,n_1 :
A_1,\ldots,n_m:A_m$.
Observe that the as yet unspecified part of a context is elided in
determining weak well-typing.  
Finally, the property of being weakly well-typed is extended to
arbitrary formulas.
This extension works in the obvious way with respect to the
propositional connectives.
Quantifiers over contexts are elided and quantification over an
object-level variable requires that it be possible to associate a
type with the quantified variable so that the body of the expression
is weakly well-typed in the extended assignment to the (potentially)
free variables.

A formula with no free variables is said to be well-formed if it is
well-typed under the empty assignment to free variables.
The definition of well-formedness also applies to sequents.
Suppose that the sequent under consideration is
$\seq{\mathcal{C}}{\Psi}{\Delta}{F}$.
In the first step, we replace each free variable $\Gamma$ in $\Delta$
and $F$ that ranges over contexts and for which a declaration of the
form $\Gamma : id[(b_1),\ldots,(b_n)]$ appears in
$\mathcal{C}$ by the sequence $b_1,\ldots,b_n$.
Let the result of this replacement be $\Delta'$ and $F'$ respectively.
The sequent is then well-formed if there is an assignment of
types in the simplified vocabulary to the variables in $\Psi$ under
which $F'$ and each member of $\Delta'$ is weakly well-typed and, for
each declaration of the form $\Gamma : id[(b_1),\ldots,(b_n)]$ that
appears in $\mathcal{C}$, the context $b_1,\ldots,b_n$ is weakly
valid.

We will assume that all the closed formulas and sequents we consider are
well-formed in the sense described above; one of the
requirements for our proof rules will, in fact, be that they preserve
the well-formedness of sequents. 
This weak well-typing property suffices to guarantee that
$\lambda$-normal forms exist for all the expressions that will occur
in our proofs~\cite{harper07jfp}.

\subsection{The Treatment of Case Analysis}

The most intricate issue when searching for a proof for sequents in
our logic is how to make use of an atomic assumption formula.
Such a formula has the form $\{ G \vdash M : A \}$ in which we can
assume that $M$ and $A$ are in $\lambda$-normal form and $M$ has a
structure that is determined by that of $A$.
Since this formula appears as an assumption, we need to consider the
different ways in which the judgment it represents in LF might have
been derived and show that the corresponding refinement of the sequent
in each case must have a proof.
If $A$ is a non-atomic type, then there would be only one way in which
the judgment could have been derived and the refinement that must be
made to the goal sequent is obvious.
Thus, the only case that we need to consider carefully is when the
sequent has the form
$\seq{\mathcal{C}}{\Psi}{\Delta\uplus\big\{\{G \vdash {M : a\app
N_1\ldots\app N_m}\}\big\}}{F}.$ 

From a little reflection, it is clear that the analysis of the atomic
formula in question should be driven by the type-level constant $a$
and by the declarations in $G$ and in the signature $\Sigma$ that
allow us to construct terms that have a type with $a$ as head.
Let $u$ be a constant in $\Sigma$ or a nominal constant
determined by $G$ that meets this condition and that the type
associated with it is
$\typedpi{x_1}{A_1}{\ldots\typedpi{x_n}{A_n}{(a\app B_1\app\ldots\app B_m)}}$. 
Each such $u$ gives rise to a collection of cases in the case analysis
that is determined as follows.
We first introduce new eigenvariables $X_1,\ldots,X_n$ and
raise them over all the nominal constants determined by the context
$G$ to yield the terms $t_1,\ldots, t_n$; $t_i$ in this sequence would
have the form $(X_i\app n_1\app \ldots\app n_\ell)$ where $n_1,\ldots,n_\ell$
is a listing of the relevant collection of nominal constants.
We then consider the unifiers for the set
\[\{\langle M, (u\app t_1\app \ldots \app t_n)\rangle, \langle N_1,
B_1[t_1/x_1,\ldots,t_n/x_n]\rangle, \ldots, \langle N_m,
B_m[t_1/x_1,\ldots,t_n/x_n] \rangle\}.\]
Each unifier $\sigma$ in this collection gives rise to the task of
ensuring that the sequent
\[
\seq{\sigma(\mathcal{C})}
    {\sigma(\Psi,X_1,\ldots,X_n)}
    {\sigma(\Delta\uplus\big\{ \{\lfprove{\Sigma}{G}{X_1:A_1}\},\ldots,
                              \{\lfprove{\Sigma}{G}{X_n:\subs{A_n}{X_1/x_1,\ldots,X_n/x_n}}\}\big\} )}{\sigma(F)}
\]
is derivable; applying a substitution to the partially elaborated contexts 
and the set of assumption formulas here simply amounts to applying it
to each term appearing therein and $\sigma(\Psi)$ denotes the
eigenvariable context obtained by removing variables in the domain of
$\sigma$ and adding those in its range.
It is possible, of course, that no unifiers exist for the set
in question.
If this is true, $u$ would not be relevant to the typing derivation being
considered and no further analysis is needed in this case. 

The presentation above has blurred one important detail.
This analysis described therein is based implicitly on $M$ and $(a\app
N_1\app\ldots\app N_m)$ being, respectively, a well-formed LF term and
type. 
However, the only assurance we have from the form of sequents being
considered in the proof system is that these expressions are
weakly well-formed.
In light of this, when considering unifiers for the set
\[\{\langle M, (u\app t_1\app \ldots \app t_n)\rangle, \langle N_1,
B_1[t_1/x_1,\ldots,t_n/x_n]\rangle, \ldots, \langle N_m,
B_m[t_1/x_1,\ldots,t_n/x_n] \rangle\}\]
we ignore possible dependencies in types, i.e. we reduce the problem
to unification in the simply typed $\lambda$-calculus.
The collection of unifiers that is obtained under this more liberal
form of typing will cover all the possibilities when the expressions
are restricted to being well-formed in LF and thus the resulting case
analysis will be conservative.\footnote{In the implementation, case
  analysis will be further limited to the situations in which
  higher-order pattern unification is applicable. The corresponding
  collection of terms admits the notion of most general
  unifiers. Thus, the liberalization of type structure we will
  sometimes cause us to consider a case when dependencies in typing
  might in fact rule it out.} 
Note also that the unifiers obtained in this way will be weakly
well-typed and the application of such unifiers to weakly well-typed
expressions preserves weak well-typing.
Using these facts, we can show that the premises that we have
described for a particular branch in case analysis satisfy the
restrictions we have placed on sequents in our proof system. 

To complete the picture related to the treatment of case analysis, we
need to explain how a context $G$ determines the nominal constants
relevant to such an analysis.
These constants could appear explicitly in $G$. If $G$ is of the form
$\Gamma, n_1:A_1,\ldots, n_k:A_k$ and $\Gamma$ is declared to have the
type $id[(b_1),\ldots,(b_p)]$, then they could also be constants declared
in one of the ``blocks'' $b_i$.
Finally, they could belong to one of the implicit parts in the
schematic view of $\Gamma$ that we described earlier in the section.
In this last case, we must make the relevant block explicit in the
type declaration for $\Gamma$ before we can use it.
This is done as follows.
Suppose that the block definition in the context schema corresponding
to $id$ that we want to expand has the form
$(x_1:A_1,\ldots,x_n:A_n).[y_1:B_1,\ldots,y_m:B_m]$.
Further, suppose that we want to introduce it right after block $b_i$
in the sequence of blocks in the type associated with $\Gamma$.
We generate new eigenvariables $Z_1,\ldots,Z_n$ and create terms
$s_1,\ldots,s_n$ that correspond to raising these variables over the
nominal constants in the blocks $b_1,\ldots,b_i$.
Then, using fresh nominal constants $n_1,\ldots,n_m$, we add the block 
$(n_1:B_1[s_1/x_1,\ldots,s_n/x_n],\ldots,n_m:B_m[s_1/x_1,\ldots,s_n/x_n])$
at the desired location in the type of $\Gamma$.
To encode the required typing constraints we augment the collection of
assumption formulas for the sequent with the formula
$\{G \vdash {s_i: A_i[s_1/x_1,\ldots,s_{i-1}/x_{i-1}]}\}$ for
each $s_i$.
The refinement of the context also introduces new nominal constants
that can be used in terms in the sequent.
To encode this possibility, we replace all the eigenvariables which appear
in the sequent only within the scope of the newly explicit block
with terms that raise them over the newly introduced nominal
constants.
At this stage, the consideration of this case proceeds as previously
described.

It may appear from the description above that treating an atomic
assumption can be complex in that it might introduce many cases to
consider.
In practice we expect this not to be the situation because the parts
of contexts that need to be made explicit in proofs will be quite
limited.

\section{An Example}
\label{sec:example}

We illustrate the logic sketched in the previous section by
considering a proof of the formula 
\[ \Pi_{tyctx}\Gamma.\forall E.\forall {\it Ty}_1.\forall {\it Ty}_2.\forall D_1.\forall D_2.
\{\lfder{\Gamma}{D_1:\of~E~{\it Ty}_1}\}\supset
\{\lfder{\Gamma}{D_2:\of~E~{\it Ty}_2}\}\supset
\exists D_3.\{\lfder{\cdot}{D_3:eq~{\it Ty}_1~{\it Ty}_2}\} \]
presented in Section~\ref{sec:formulas}. Such a proof will need
induction that is built into our logic in a manner similar to that in
Abella~\cite{baelde14jfr}: we add the formula that we want to prove to
the assumption set of the sequent, marking one of the antecedents in
it with the understanding that this antecedent can be matched only by
some formula that is obtained by case analysis on the corresponding
assumption formula.
In this case, we will use induction with respect to the first antecedent,
i.e. $\{\lfder{\Gamma}{D_1:\of~E~{\it Ty}_1}\}$. 

Starting from the initial state $\seq{\cdot}{\cdot}{\cdot}{F}$, where $F$ 
represents the formula above, using induction and obvious
introduction rules will leave us wanting to prove the sequent
\begin{gather*}
\seq{\Gamma:tyctx[\cdot]}{E,{\it Ty}_1,{\it
Ty}_2,D_1,D_2}{\big\{IH,
\{\lfder{\Gamma}{D_1:\of~E~{\it Ty}_1}\},
\{\lfder{\Gamma}{D_2:\of~E~{\it Ty}_2}\}\big\}}{}\\
\exists D_3.\{\lfder{\cdot}{D_3:eq~{\it Ty}_1~{\it Ty}_2}\}
\end{gather*}
in which $IH$ is a version of $F$ to which the interpretation
explained above has been attached.
At this point we consider case analysis on the formula
$\{\lfder{\Gamma}{D_1:\of~E~{\it Ty}_1}\}$. 
There will be three broad categories to consider for the claimed LF
derivation: it was concluded using the $\ofapp$ or $\oflam$
constructors drawn from the signature or by using something from the
context. 
The first two cases will result in the structure of the simply-typed
term (represented by $E$) being elaborated and this will constrain
the cases in a subsequent case analysis on the premise deriving from
the second antecedent in $F$.
The proof can then be concluded by using the induction
hypothesis.\footnote{In the case of $\oflam$, to use the induction
hypothesis, we would need to recognize also that items introduced into
the context satisfy the context schema corresponding to $tyctx$.}

Turning to the third case, we see that the context variable $\Gamma$
is currently completely unspecified.
Thus the only possibility for using something from the context would
arise if we employed the definition of $tyctx$ to introduce the block 
$(n_1:tm,n_2:of~n_1~T)$ explicitly into $\Gamma$.
Specifically, this case will lead to the type of $\Gamma$ in
$\mathcal{C}$ being refined to $tyctx[(n_1:tm,n_2:of~n_1~T)]$ and the 
eigenvariables $E,{\it Ty}_1,{\it Ty}_2,D_1,$ and $D_2$ being raised
over $n_1$ and $n_2$. 
Further, $\{\lfder{\Gamma}{T:ty}\}$ will be added as an
assumption formula to the sequent. 
Once this is done the LF judgment we are considering will be of the following form 
$\{\lfder{D_1~n_1~n_2:of~(E~n_1~n_2)~({\it Ty}_1~n_1~n_2)}\}$.
Only the second declaration in the introduced block can ``match'' with
this case.
Using it will result in the substitution of $\lambda x. \lambda
y. n_1$ for $E$, $\lambda x.\lambda y. n_2$ for $D_1$ and $\lambda
x.\lambda y. T$ for $ty$. 
Our state at the end of this will be the sequent
\begin{gather*}
\seq{\Gamma:id[(n_1:tm,n_2:of~n_1~T)]}{{\it
Ty}_2,D_2,T}{\big\{IH,\{\lfder{\Gamma}{T:ty}\},
\{\lfder{\Gamma}{D_2~n_1~n_2:of~n_1~({\it Ty}_2 \app n_1 \app n_2)}\}\big\}}{} 
\\
\exists D_3.\{\lfder{\Gamma}{D_3:eq \app T \app ({\it Ty}_2\app n_1\app
n_2)}\}.
\end{gather*}
At this stage, we may try a case analysis on the assumption
formula $\{\lfder{\Gamma}{D_2\app n_1\app n_2:of\app n_1\app ({\it
Ty}_2 \app n_1 \app n_2)}\}$.
The only possibility that would work here would correspond to using the second
declaration in the only block in the elaborated form for $\Gamma$.
In this case ${\it Ty}_2$ will be set to $\lambda x. \lambda y. T$.
The resulting sequent has an easy proof. 
\section{Further Developments to the Work}
\label{sec:furtherwork}

This paper has described preliminary work on the development of a
logic for reasoning about LF specifications.
There are three specific directions in which we plan to continue this
work to bring the project to a more complete state.

The first direction involves that of establishing meta-theoretic
properties for the logic.
The formulas that are permitted by the logic are fairly unrestricted.
However, to obtain a handle on mechanical reasoning, we limited
formulas and sequents that would be considered by the proof system
based on a notion of weak well-typing.
We consequently do not expect this system to be complete.
However, the logic must be sound in order to be useful.
This is a property that we have had in mind in designing the rules but
the design considerations need to now be translated into an actual proof.
The rule that needs most careful examination in this regard
is the one for case analysis, i.e. the only rule we have
discussed in some detail here.
The key here would be to show that the analysis embodied in the rule
covers all possible ways in which an LF derivation encoded in the formula
being examined could have been concluded.
We believe this to be true---the analysis may actually cover ``more
cases'' than needed because of the use of unification over weakly
well-typed terms---but a proof of this fact still has to be completed.
Soundness will immediately yield consistency for the proof system
since there are obviously formulas in the logic that are
not valid.

The remaining directions of work concern the implementation of our
logic and understanding its usefulness by experimenting with
applications.
The former aspect will stimulate the latter: working with sufficiently
large and, hence, meaningful examples is impacted significantly by the
availability of a mechanical tool.
It is therefore a matter of immediate concern and in fact one that is
being addressed by our current research.

\section{Conclusion}
\label{sec:conclusion}

This paper has presented work towards defining a logic for reasoning about
LF specifications.
While this work is not unique in terms of its goal, it does differ
from other developments in terms of how it is attempting to achieve
its objective.
The Twelf system~\cite{Pfenning02guide} provides a means for reasoning
about specifications through checking the totality of relations encoding
the desired meta-theoretic property.
In contrast, we are proposing a meta-logic for reasoning about LF 
specifications and so we will be able to make explicit the proofs of
the relevant properties. 
The logic $M_2^+$~\cite{schurmann00phd} has also been developed for reasoning 
about LF specifications.
In this logic the context is not made explicit, and so it is not capable of
expressing properties which, for example, may involve LF judgments in
distinct contexts.
The ways of stating properties in $M_2^+$ and our logic are also 
distinct, and it is of future interest to understand what these
differences imply in terms of capabilities. 
Finally, there is work on the use of a translation of LF specifications 
into predicate logic which might be used to reason about these specifications
using the Abella system~\cite{southern14fsttcs}.
A drawback with this approach is that it requires the user to
understand and to work with the translation that is used. 
Our goal in this work has been to let the focus reside entirely on the
analysis of derivability in LF.

\section*{Acknowledgements}

This work has been supported by the National Science Foundation grant
CCF-1617771. Opinions, findings and conclusions or
recommendations that are manifest in this material are those of the
participants and do not necessarily reflect the views of the NSF.

\bibliographystyle{eptcs}
\bibliography{master}
\end{document}